\documentclass[journal]{IEEEtran}
\usepackage{amsmath,amsfonts}
\usepackage{array}
\usepackage{textcomp}
\usepackage{stfloats}
\usepackage{url}
\usepackage{verbatim}
\usepackage{graphicx}
\usepackage{color}
\usepackage{xcolor}
\usepackage{pifont}
\usepackage{enumitem}

\usepackage{mdframed}

\usepackage{mathtools} %splifrac

\usepackage{threeparttable}

\usepackage{array} % 加载 array 包（尽管在这个例子中不是必需的，但通常用于表格定制）  

\definecolor{mjColorRevision}{rgb}{0.0,0.0,1.0}
\definecolor{mjColorComment}{rgb}{1.0, 0.0, 0.0}

\definecolor{boxtitlecolor}{gray}{0.7} % ??boxbackground???60%???
\definecolor{boxbgcolor}{gray}{0.96} % ??boxbackground???60%???

\definecolor{hlcllink}{rgb}{0.0, 0.0,1.0}
\usepackage{hyperref}  %%

\definecolor{blue-green}{rgb}{0.0, 0.87, 0.87}
\definecolor{hlgreen}{rgb}{0.0,0.7,0.0}

\newcommand{\xmark}{\ding{55}}
\newcommand{\lmark}{\ding{51}}

\newcommand{\hlxitem}{\textcolor{red}{\xmark}}
\newcommand{\hllitem}{\textcolor{hlgreen}{\lmark}}

\definecolor{mjColorComment}{rgb}{1.0, 0.0, 0.0}
\definecolor{mjColorCommentNC}{rgb}{0.9, 0.5, 0.2}

\definecolor{mjColorInternalChat}{rgb}{0.0,0.0,1.0}
\definecolor{mjColorInternalChatR}{rgb}{0.0,0.0,0.0}

\usepackage{cite}

\usepackage[linesnumbered,ruled]{algorithm2e} %hlma
\usepackage{multirow} %hlma
\usepackage{cellspace}
\usepackage{makecell}
\usepackage{amssymb}
\usepackage{cases}

\hypersetup{
	colorlinks=true, % 启用彩色链接
	linkcolor=hlcllink,   % 内部链接（如章节、引用等）颜色
	filecolor=hlcllink,   % 文件链接颜色
	urlcolor=hlcllink,    % URL 链接颜色
	citecolor=hlcllink    % 引用链接颜色
}

\usepackage{theorem}
\usepackage[tight,footnotesize]{subfigure}

\def\myproof{\noindent{{\textbf{Proof:}}}} %hlma
\ifdefined\myproof
\else
\def\myproof{\proof}

\fi

\newtheorem{proposition}{Proposition}
\newtheorem{corollary}{Corollary}

\newtheorem{remark}{Remark}
\newtheorem{proof}{Proof}

\definecolor{hlgrey}{rgb}{0.2,0.2,0.4}
\definecolor{hlgreen}{rgb}{0.0,0.4,0.0}
\definecolor{hlyellow}{rgb}{0.5,0.6,0.0}
\definecolor{hlcolortemp}{rgb}{1.0,0.0,0.8}

\ifdefined\myproof
\else
\def\myproof{\proof}

\fi

\makeatletter

\newcommand{\Rmnum}[1]{\expandafter\@slowromancap\romannumeral #1@}
\makeatother

\begin{document}

\ifCLASSOPTIONcaptionsoff
\newpage
\fi

\title{
	% Effective  Range of Coils-Based Through-The-Earth Magnetic Induction Communication
	% \color{blue} Optimizing Through-The-Earth Magnetic Induction Communication: Effective Range and Optimal Frequency Analyses
	Effective Range and Optimal Frequency of Through-the-Earth Magnetic Induction Communication
}

\author{IEEE Publication Technology,~\IEEEmembership{Staff,~IEEE,}
	% <-this % stops a space
	\thanks{This paper was produced by the IEEE Publication Technology Group. They are in Piscataway, NJ.}% <-this % stops a space
	\thanks{Manuscript received April 19, 2021; revised August 16, 2021.}}

\author{Honglei~Ma,
	Erwu~Liu, \IEEEmembership{Senior Member,~IEEE,}
	Wei~Ni,~\IEEEmembership{Fellow, IEEE,}
	Jun~Zhu,\\
	Zhijun~Fang, \IEEEmembership{Senior Member,~IEEE,}
	Rui~Wang, \IEEEmembership{Senior Member,~IEEE,}
	Yongbin~Gao, and 
	Xinyu~Qu
	\thanks{This work was sponsored by the National Natural Science Foundation of China under Grant 62271352 and the Fundamental Research Funds for the Central Universities.}
	\thanks{H. Ma, Z. Fang, and Y. Gao are with the
		School of Electronic and Electrical Engineering, Shanghai University of Engineering Science, Shanghai 201620, China. (e-mail: holyma@yeah.net,   Zjfang@gmail.com, gaoyongbin@sues.edu.cn). }% <-this % stops a space
	\thanks{E. Liu is with the Department of Ophthalmology, Tongji Hospital, School of Medicine, Tongji University, Shanghai 200070, China (e-mail: erwu.liu@ieee.org). }
	\thanks{R. Wang, and X.~Qu are with the School of Electronics and Information Engineering, Tongji University, Shanghai 201804, China. R. Wang is also  with the Shanghai Institute of Intelligent Science and Technology, and  National College of Elite Engineers, Tongji University(e-mail:  ruiwang@tongji.edu.cn, xinyuqu@tongji.edu.cn).}% <-this % stops a space
	%	\thanks{W. Ni is with ICT Centre, CSIRO, Marsfield, NSW 2122, Australia. (e-mail: wei.ni@ieee.org)}
	\thanks{W. Ni is with the School of Engineering, Edith Cowan University, Perth, WA 6027, Australia (e-mail: wei.ni@ieee.org).}
	\thanks{J. Zhu is  with  the Shanghai Institute of Microsystem and Information Technology, Chinese Academy of Sciences, Shanghai, China (e-mail: zhuj@mail.sim.ac.cn). }
	\thanks{ (\emph{Corresponding authors: E. Liu})}
	\thanks{DOI: 10.1109/TVT.2026.3721188 }

	% <-this % stops a space
	%	\vspace{-1.6em}
}

%\markboth{Accepted by IEEE Transactions on Vehicular Technology for publication, DOI: 10.1109/TVT.2026.3721188}

\markboth{Accepted by IEEE Transactions on Vehicular Technology for publication, DOI: 10.1109/TVT.2026.3721188}%
{Shell \MakeLowercase{\textit{\emph{et al.}}}:Fast-Fading Channel and Power  Optimization of the Magnetic Inductive Cellular Network}


\maketitle

\begin{abstract}
	
	Magnetic induction communication (MIC) is a promising technology for through-the-earth (TTE) communication.
	Previous studies on the MIC range have often overlooked the impact of eddy losses caused by underground materials. For TTE MIC, significant eddy losses complicate the analysis of the effective MIC range, which is vital for optimizing performance but has never been addressed in the literature. 
	Accounting for the conductivity and permittivity of the underground medium, this paper derives  the effective MIC range in TTE MIC, along with a closed-from expression that  predicts the optimal carrier frequency to maximize this range.
	Finite element simulations validate the analysis, demonstrating that the optimal carrier frequency can significantly enhance the MIC range. It is also revealed that optimizing the antenna radius is effective in extending the MIC range for TTE and vehicle MIC  applications.
\end{abstract}


\section{Introduction} \label{sect_introduction}

Through-the-earth (TTE) communication (TTEC), also known as deep and long-distance underground communication, has a communication range exceeding tens of meters. It constitutes a foundational communication link for the underground Internet of Vehicles (IOV).   Since electromagnetic (EM) waves suffer significant path losses in deep underground scenarios,  TTEC poses a bottleneck but is of practical significance for underground applications, such as subterranean tunnels~\cite{Ma2019Antenna}, mining~\cite{Dong2020Velocity}, robotics\cite{Sun2024Improved}, subway~\cite{Ko2024Field}, and resource exploration. 
Magnetic induction communication~(MIC), primarily in the near-field (NF) regime~\cite{Liu2024Magnetic}, offers a good option for underground communications~\cite{Guo2017Multiple,Sun2010Magnetic,Guo2021Joint,Sun2013Increasing}. Using 8 m transmit antenna, our team  achieved a vertical transmit distance of 310 m in a coal mine environment in Datong, China\cite{zhang2014cooperative, Zhang2017Connectivity}.

%Unlike EM waves,
The effective MIC range is the most important performance metric in TTEC since such communication is a bottleneck of deep underground applications.  Early studies on MIC emphasized on the MIC distance. In~\cite{Sun2010Magnetic}, researchers simulated the path loss and bit error rate (BER) with respect to (w.r.t.) the distance of the MIC and magnetic induction (MI) waveguide. In~\cite{kisseleff2013channel}, researchers discussed the capacity w.r.t. the distance of the MIC and MI waveguide. The authors of~\cite{Li2022Optimal} evaluated the meta-material MIC distance by simulations.

As these studies \cite{Sun2010Magnetic, kisseleff2013channel,Li2022Optimal} were based on short- and mid-range MIC, the eddy current effect caused by underground materials has often been overlooked. The channel power gain is $H_{SD}(d_{}) \simeq c d^{-6}$, where $d_{}$ is the communication distance and $c$ is a constant. 
This makes the evaluation of the MIC range straightforward. However, in TTE scenarios, the eddy currents generated in the underground medium are typically non-negligible and complicate the function  $H_{SD}(d_{})$,  especially when accounting for the permittivity of underground materials. Additionally, these studies using the frequencies of 1--30 MHz offer a shorter near-field range.

For TTE environments, particularly those involving vehicle MIC (VMIC) where deploying MI waveguide systems~\cite{Sun2010Magnetic,kisseleff2013channel} is challenging, methods of very low frequency (VLF) signals and larger antennas (VLF-LA)  were developed in~\cite{zhang2014cooperative,Ma2019Effect,Ma2019Antenna,Ma2024Fast}, achieving distances of 60--310 m for TTEC.  Some studies, e.g.,~\cite{zhang2014cooperative}, have primarily determined the TTE MIC range through numerical methods, and derived a closed-form expression in \cite{Zhou2017Maximum}. However, these studies \cite{zhang2014cooperative, Zhou2017Maximum} overlooked both  the circuit loss caused by coil resonance and the eddy loss induced by underground medium, both of which are  significantly frequency-dependent. Zhang et al. \cite{Zhang2017Connectivity} derived the expression for MI coverage  via the Lambert-W function, accounting for eddy and circuit losses in an MI network connectivity algorithm. However, for  faster convergence, they simplified the eddy model by omitting underground permittivity since introducing this parameter would significantly increase the complexity.  This  limits practical relevance.

%To enhance the communication range, studies in~\cite{Sun2010Magnetic,kisseleff2013channel} proposed the MI waveguide achieving an MIC distance of 250 m. Due to the limited free underground space, it is difficult to deploy passive relays in TTE and vehicular communication scenarios. 

Most of these studies on TTEC \cite{zhang2014cooperative,Ma2019Effect,Ma2019Antenna,Ma2024Fast} have adopted the carrier frequency of 10 kHz, which appears to be an empirical value. This  may not be suitable for various underground environments. Finding a higher carrier frequency for TTE MIC to maximize the MIC range is crucial for VLF-LA methods. It also facilitates an increase in bandwidth for VLF-LA methods. However, few  studies have addressed this issue, which motivates us to derive the optimal MIC distance and its corresponding carrier frequency for VLF-LA techniques. %The non-elementary function of the MIC range also makes it challenging to derive this frequency.

%In these works, the papers~\cite{Ma2019Antenna} and~\cite{Ma2019Effect} investigate  the achievable rate of MI TTEC, the work~\cite{Ma2020Channel} studies the fast-fading channel of the MI TTEC, and the literature~\cite{zhang2014cooperative} proposes an active relay to extend the MIC distance. Unfortunately,  researchers in~\cite{zhang2014cooperative} mainly focus on the optimization of the orientation of the antenna instead of the rules of the MIC distance.

%As the communication distance extends, the eddy current generated by magnetic field signals within underground materials escalates. This leads to a substantial decline in energy efficiency. Consequently, the effects of underground materials cannot be overlooked. Namely, the intricacies of MIC link channel gain in long-range MIC models become increasingly evident. According to~\cite{kisseleff2013channel}, the eddy current in the underground medium can be simplified as the skin depth model.  The skin depth significantly depends on the carrier frequency which can be artificially optimized. However, few previous works focus on it. To the best of our knowledge, this letter is the first paper on obtaining the optimal MIC distance through frequency optimization.
\IEEEpubidadjcol

% The contributions of this paper are summarized as follows: 1) 
In this paper, we derive a closed-form expression for  effective MIC range, overcoming the complexity induced by eddy loss and medium permittivity. Based on this expression, we analyze the optimal TTE MIC range and find that increasing the coil radius to an appropriate value is effective in extending the MIC range, whereas increasing the transmit power is not.
We also derive a proposition and a  practically applicable corollary with closed-form expression for predicting the optimal frequency in TTE MIC.
This frequency can potentially extend the MIC range by multifold, and minimally affected by the radiated field. The  advantages of our approach, compared to those in prior studies, are as summarized in Table  \ref{tbl_advantage}.

\begin{table*}[htp] \scriptsize%\footnotesize
	\centering
	\label{tbl_advantage}
		\caption{Advantages of the proposed approaches}
		\vspace{-0.8em}
		\scalebox{0.95}{
			\begin{threeparttable}
				\begin{tabular}{m{0.04\textwidth}<{\centering}|m{0.082\textwidth}<{\centering}|m{0.42\textwidth}<{\centering}|m{0.40\textwidth}<{\centering}}
					\hline
					\textbf{Refs.} & \textbf{Model type} & {\textbf{Addressed issues methods, and remaining issues$^\dagger$}} & {\textbf{Advantages of this paper  (v.s. the target ref.$^\dagger$})}  \\
					\hline \hline
					\cite{zhang2014cooperative}&Near-field, weak-coupling&\begin{itemize}[leftmargin=6pt]
						\item[\hllitem]Numerical analysis of the MIC range; lack of closed-form expression
						\item[\hlxitem]Overlooking  eddy loss and medium permittivity 
						\item[\hlxitem]Overlooking  circuit loss
						%	\vspace{-1.0em}
					\end{itemize} &\begin{itemize}[leftmargin=6pt]
						\item[\hllitem] Closed-form expressions integrating effects of eddy loss and medium permittivity
						\item[\hllitem] Proposition for optimal frequency prediction
						%	\vspace{-1.0em}
					\end{itemize}\\ \hline	
					%%%%%%%%%%%%%%%%%%%%%%%%%%%%%%%%%%%%%%%%%%%%%%%%%%%%%%%%%%%%%%%%%%%%%%%
					%%%%%%%%%%%%%%%%%%%%%%%%%%%%%%%%%%%%%%%%%%%%%%%%%%%%%%%%%%%%%%%%%%%%%%%%%%%%%%%%%%%%%%%%%%%%%%%%%%%%%%%
					\cite{Zhou2017Maximum}&Far-field&\begin{itemize}[leftmargin=6pt]
						\item[\hllitem]Closed-form expression of the MIC range
						\item[\hlxitem] Neglect of eddy loss and medium permittivity
						\item[\hlxitem] Neglect of circuit loss
						%	\vspace{-1.0em}
					\end{itemize} &\begin{itemize}[leftmargin=6pt]
						\item[\hllitem] Closed-form expression accounting for eddy loss and medium permittivity effects
						\item[\hllitem] Proposition for optimal frequency prediction
						%	\vspace{-1.0em}
					\end{itemize}\\ \hline
					%%%%%%%%%%%%%%%%%%%%%%%%%%%%%%%%%%%%%%%%%%%%%%%%%%%%%%%%%%%%%%%%%%%%%%%%%
					\cite{Zhang2017Connectivity}&Near-field, weak-coupling&\begin{itemize}[leftmargin=6pt]
						\item[\hllitem]  Closed-form expression accounting for eddy loss
						\item[\hlxitem] Neglect of medium permittivity
						%	\vspace{-1.0em}
					\end{itemize} &\begin{itemize}[leftmargin=6pt]
						\item[\hllitem]  Consideration of medium permittivity and ~33\% MIC range extension potential via MFRE
						\item[\hllitem]  Comprehensive analysis and prediction for MIC range optimization
						%	\vspace{-1.0em}
					\end{itemize}\\ 	\hline
					This paper&Near-field, weak-coupling&\begin{itemize}[leftmargin=6pt]
						\item[\hllitem] Closed-form expressions with eddy  and  permittivity effects
						\item[\hllitem] Comprehensive analysis and predicting for MIC range optimization 
						%	\vspace{-1.0em}
					\end{itemize} &-\\ 	
					\hline      
					
				\end{tabular}
				%%%%%%%%%%%%%%%%%%%%%%%%%%%%%%%%%%%%%%%%%%%%%%%%%%%%%%%%%%%%%%%%%%%%%%
				
				\begin{tablenotes}  
					\footnotesize  
					\item[$\dagger$] {The markers `\hllitem' and `\hlxitem'  represent addressed and  remaining issues, respectively.}
				\end{tablenotes}  
			\end{threeparttable}
		} 
	  
	\vspace{-1.7em}
\end{table*}

%\vspace{-0.5em}

\section{Analysis of Effective MIC Range}\label{sect_range}

The effective MIC range constitutes the foremost performance metric for TTE applications. As the eddy current is often  ignored in previous studies on the MI distance\cite{Sun2010Magnetic}, obtaining the MIC range may initially seem straightforward. However, as the underground permittivity cannot be ignored in the TTEC,  the analysis of the effective MIC range becomes non-trivial.

The effective communication range is the largest distance between the transmitting and receiving nodes when data can be transmitted correctly~\cite{Rappaport1996Wireless}. Such range primarily depends on the channel fadings. Upper-layer protocols, such as modulation and error correction coding, bring additional losses in the MIC range. These losses can be equivalently represented as an overall signal-to-noise ratio (SNR) threshold, denoted by~$\Upsilon_{\rm th}$.

The effective MIC range is defined as the distance $d_{}$ between the receiving and transmitting nodes when the SNR of the receiving node $\Upsilon_{SD}$ is below a certain threshold $\Upsilon_{\rm th}$, i.e., $\Upsilon_{SD}(d_{})$$=$$\Upsilon_{\rm th}$, and $\Upsilon_{\rm th}$ represents the overall equivalent SNR loss, including the loss in upper-layer protocols.

\begin{figure}[t]
	\vspace{0em}
	\centering{\includegraphics[width=70mm]{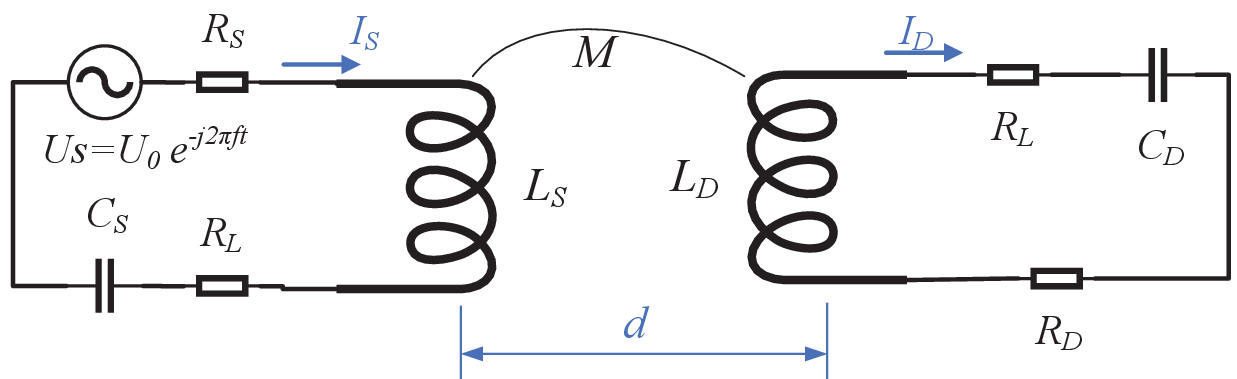}}
	\caption{Equivalent circuit model for the MIC link $S$$\rightarrow$$D$. $S$ and $D$ denote the transmit  and  receive nodes, respectively.  $L_S$ is a transmit coil with the inductance $L_S$, the radius $a_S$ and the number of turns $n_S$.  $L_D$ is a receive coil with the inductance $L_D$, the radius $a_D$ and the number of turns $n_D$. $C_S$ and $C_D$ are the matching capacities.
		\vspace{-0.0em}
	}\label{fig_P2PCircuit}
	\vspace{-1.5em}
\end{figure}

According to~\cite{Ma2019Antenna}, the MIC link, $S$--$D$, can be modeled as shown in Fig.~\ref{fig_P2PCircuit}, where the receive SNR at node $D$ is $\Upsilon_{SD} (d_{})$$=$$\tfrac{P_S  H_{SD} (d_{})}{P_N}$$=$$h_0 f^2  |M_{} (d_{})|^2$.
Here, $P_S$ is the transmitting power spectral density (PSD) of the transmit node $S$, ${P_N}$ is the noise PSD  at the receive node $D$, $H_{SD}$ is the channel power gain of the link $S$--$D$, and $d_{}$ is the distance between the nodes $S$ and $D$, 
$h_0$$=$$\left|\frac{(2\pi)^2 R_L}{ Z_D^2 Z_S}\right|$; $Z_S$$=$$j 2\pi f {L_{S}}$$+$$\frac{1}{j2\pi f C_{S}}$$+$$R_{S}$$+$$R_L$ and  $Z_D$$=$$j 2\pi f {L_{D}}$$+$$\frac{1}{j2\pi f C_{D}}$$+$$R_{D}$$+$$R_L$ denote the overall circuit impedances of the transmit and receive coils, respectively,   $M$ is the  mutual inductance between coils $S$ and $D$, For the mixed-field (MF) and based on~\cite{Ma2026Through}, $M$ can be deduced as
\begin{equation}\label{eqn_msdmix}
	\begin{aligned}
		M_{}(d) &= (\pi \mu a_S^2 a_D^2 n_S n_D) \tfrac{ 1}{4d^3} J_{SD} e^{-jk_0 d_{}}, \\
		J_{SD} &= [2 \cos\theta_{S} \cos\theta_D\left(1\!+\!jk_0 d_{} \right)
		\\ &+ \sin\theta_{S}\sin\theta_D   \left(1 + jk_0 d_{} -(k_0 d_{})^2\right)],
	\end{aligned}
\end{equation}
where $\theta_S$ and $\theta_D$ represent the  orientations of  antennas $S$ and  $D$, respectively,  $k_0$$=$$2\pi f \sqrt{\mu ( \varepsilon\!+\!j \frac{\sigma}{2\pi f} )}$ is the propagation constant. Here, $\mu_{}$, $\sigma$ and $\epsilon$ are the permeability, conductivity and permittivity of the underground medium, respectively. Thus, the effective MIC range can be obtained by solving the MF range equation (MFRE):
\begin{equation}\label{eqn_dsdeq}
	h_0 f^2 | M_{} (d_{})|^2 = {\Upsilon_{\rm th}}.
\end{equation}	
Let $\beta$$=$$\operatorname{Re}\{k_0\}$$=$$2\pi f \sqrt{\tfrac{\mu}{2}[\sqrt{\varepsilon^2\!+\!(\tfrac{\sigma}{2\pi f})^2}]\!+\!\varepsilon}$ and  $\alpha$$=$$\operatorname{Im}\{k_0\}$$=$$2\pi f \sqrt{\tfrac{\mu}{2}[ \sqrt{\varepsilon^2\!+\!(\tfrac{\sigma}{2\pi f})^2}\!-\!\varepsilon]}$ which correspond to the phase constant and attenuation constant, respectively. Re$\{\cdot\}$ and Im$\{\cdot\}$ are the real and imaginary operators, respectively. 
The factor $|e^{-jk_0 d}|$$=$$e^{-\alpha d}$ is called eddy gain. 
We adopt $d$$\leq$$\frac{0.1}{\beta}$  for the near-field region, $\frac{0.1}{\beta}$$<d$$<$$\frac{10}{\beta}$ for the Fresnel region, and $d$$\geq$$\frac{10}{\beta}$ for the far-field region.  

It is difficult to obtain the closed-form expressions of MFRE solutions. Nevertheless,  as MIC typically operates in the near-field region\cite{zhang2014cooperative, Sun2010Magnetic, kisseleff2013channel,Ma2026Through}, we focus on the near-field range  $d$$\leq$$\frac{0.1}{\beta}$.  In this region, $J_{SD}$$\simeq$$2\cos \theta_D \cos \theta_S$$+$$\sin \theta_S \sin \theta_D$ is called the misalignment factor,  and we have
\begin{equation}\label{eqn_msd0}
	\begin{aligned}
		|M(d)|\!=\! |M_{\rm NF}(d)|\!= \!(\pi \mu a_S^2 a_D^2 n_S n_D) \tfrac{ J_{SD}(\theta_S, \theta_D)}{4d_{}^3}e^{-\alpha d}.\\
	\end{aligned}
\end{equation}
After Substituting \eqref{eqn_msd0}   into~\eqref{eqn_dsdeq}, we have
\begin{equation}\label{eqn_eqrsd0}
	\begin{aligned}
		\tfrac{P_S}{P_N}  \mathfrak{C}_{SD}  J^2_{SD} \left({d_{}^3 e^{\alpha d} }\right)^{-2} &= \Upsilon_{\rm th},
	\end{aligned}
\end{equation}
where $\mathfrak{C}_{SD}$$=$$\frac{(\pi \mu a_S^2 a_D^2 n_S n_D)^2}{4} h_0 f^2$. Let $\Upsilon_S$$=$$\frac{P_S}{P_N}  \mathfrak{C}_{SD}  J^2_{SD}$,
and~\eqref{eqn_eqrsd0} is further arranged as  $\tfrac{\alpha d_{}}{3} e^\frac{\alpha d_{}}{3}$$=$$ \tfrac{\alpha}{3}\left(\tfrac{\Upsilon_S}{\Upsilon_{\rm th}}\right)^\frac{1}{6}$.
Exploiting the fact that the inverse function of $f(x)=\left(x e^x\right)$ is a Lambert-W function $\mathbb{W}(x)$, we obtain the solution to~\eqref{eqn_eqrsd0} w.r.t. $d_{}$, i.e., the effective MIC range, as given by
\begin{equation}\label{eqn_dsd}
	\begin{aligned}
		d_{}  &= \tfrac{3}{\alpha} \mathbb{W}\left( \tfrac{\alpha}{3}\left(\tfrac{\Upsilon_S}{\Upsilon_{\rm th}} \right)^\frac{1}{6} \right)\\
		&= \tfrac{3 \mathbb{W}\left( \frac{\sqrt{\mu}}{3} \sqrt{\pi  f \sqrt{4\pi^2 \epsilon^2 f^2 + \sigma^2} - 2\pi^2 \epsilon f^2}  \left( \frac{\Upsilon_{S}}{\Upsilon_{\rm th}}  \right)^{\frac{1}{6}} \right)}{\sqrt{\mu}\sqrt{\pi  f \sqrt{4\pi^2 \epsilon^2 f^2 + \sigma^2} - 2\pi^2 \epsilon f^2}}, \\
	\end{aligned}
\end{equation}
subject to the near-field upper bound
\begin{equation}\label{eqn_upbd}
	\begin{aligned}
		d&\leq\tfrac{0.1}{\beta}=\tfrac{0.1}{2\pi f \sqrt{\tfrac{\mu}{2}[\varepsilon + \sqrt{\varepsilon^2 + (\tfrac{\sigma}{2\pi f})^2}]}}.
	\end{aligned}
\end{equation}

\section{Optimal effective MIC Range} \label{sect_opt}

In this section, we derive the optimal carrier frequency aimed at enhancing the MIC range, along with an analysis of the impact of material conductivity on the MIC range.

Due to the limited free space and dynamic nature  of TTE and IoV scenarios, the use of waveguides~\cite{Jenkins2023Wearable} and active relay techniques~\cite{Zhang2024Cooperative}  becomes challenging. Compared to the general MIC, there are limited approaches to extending the TTEC and VMIC range. Directly optimizing the effective range 
% of point-to-point 
is essential for TTEC. Since the Lambert-W function is a monotonically increasing function, from~\eqref{eqn_dsd}, increasing the misalignment factor $J_{SD}$, transmit power $P_S$, and total areas of the transmit and receive coils can enhance the MIC range.  The impacts of the carrier frequency $f$ and underground materials on the MIC range require further exploration. Additionally, adjusting the coil size poses practical challenges in device deployment and mobile MIC  in deep underground environments. Therefore,  optimizing the carrier frequency $f$ is the most feasible way to extend the MIC range. Moreover, adopting a higher frequency facilitates an increase in data rate.
% , from a practical perspective.

\vspace{-1em}

\subsection{ Optimal Frequency}\label{sectsub_of}

Although adjusting the carrier frequency in a wireless system is relatively straightforward to implement,  
it is observed from~\eqref{eqn_dsd} that the function $d_{}(f)$ is non-convex due to the Lambert-W function and the permittivity effect. Thus, obtaining the optimal frequency seems to be challenging. 

Using the properties of the Lambert-W function, we obtain the following proposition on the optimal carrier frequency for achieving maximal  MIC distance.

\begin{proposition}\label{prop_fstar}
	For the objective function $d^*(f)$, let  
	\begin{subequations}\label{eqn_f0star11}
		\begin{align}
			\Upsilon^*_{\backslash f} &\triangleq \frac{P_S}{P_N} h_0 (\mu\pi n_S n_D a_S^2 a_D^2/4)^2 J^2_{SD} (\theta_S, \theta_D) , \label{eqn_snrnof}\\ 
			f_0^* &\triangleq 2^{6/5} \mathrm{e}^{4/5}  \left(\tfrac{\Upsilon^*_{\backslash f}}{\Upsilon_{\rm th}}\right)^{-1/5} \cdot (\pi\mu\sigma)^{-3/5}.	\label{eqn_f0star1}
		\end{align}
	\end{subequations}
	There exists one or two  locally optimal carrier frequencies: $f^*$ (the first stationary point) and $\infty$ (the right endpoint, which has no practical physical meaning), where the stationary point
	\begin{equation}\label{eqn_fstar}
		\begin{aligned}
			f^* &\geq f_0^*,   \\
		\end{aligned}
	\end{equation}
	with equality holding if and only if $\epsilon f=0$. In this case, $f^*$ is the unique optimal solution.
\end{proposition}

\begin{myproof}
	Please refer to Appendix \ref{sect_proof}.
\end{myproof}

Proposition \ref{prop_fstar} gives the lower bound of the optimal frequency.  This result may differ from the conventional preference  for lower frequencies when extending the TTE MIC range.  Proposition \ref{prop_fstar}  also implies that when  $f^*$ does not exist, $d_{} (f)$  increases monotonically with $f$. Notably, higher frequency can also facilitate boosting data rates.

For the near-field bound, we have $d_{\rm NF}$$=$$\frac{0.1}{\beta}$$\leq$$ \frac{0.1}{\sqrt{\pi\mu\sigma f_{\rm NF}}}$ (i.e.,  $\pi\mu\sigma \leq f^{-1}_{\rm NF}/100d^2_{\rm NF}$), where $f_{\rm NF}$ is the frequency at the near-field upper bound  $d_{\rm NF}$. When the transmit power is not excessively high (e.g., $P_S<10^6$ W), we have
\begin{equation}\label{eqn_f0leqfnf}
	f^*_0 \geq 2^{6/5}  \mathrm{e}^{4/5} \left( \tfrac{\Upsilon^*_{\backslash f}}{\Upsilon_{\rm th}}\right) ^{\frac{-1}{5}} (10^\frac{6}{5}d^{\frac{6}{5}}_{\rm NF}) f^\frac{3}{5}_{\rm NF} > f_{\rm NF}.
\end{equation}
Similarly, for the far-field  ($d$$\geq$$\frac{10}{\beta}$), if $P_S$ is not too low (e.g., $P_S$$<$1 W), the inequality $f_0^*$$\leq$$f_{\text{FF}}$ holds, where $f_{\text{FF}}$ is the frequency at the far-field lower bound. Therefore, it can be considered that $f^*_0$ lies within the Fresnel region (i.e., $f_{\rm NF}$$<$$f^*_0$$<$$f_{\rm FF}$).

As $\lim\limits_{\epsilon f\rightarrow0}\frac{\sigma}{2\pi \epsilon f}$$=$$\infty$, i.e., $f$$\ll$$\frac{\sigma}{2 \pi  \epsilon}$, the  attenuation constant $\alpha  \simeq {\pi f \mu_{} \sigma_{}}$ holds.
	From Proposition \ref{prop_fstar},   a corollary for more practical applications can be derived as follows.
	
	\begin{corollary}\label{cor_fstar}
		For the objective function  $d^*(f)$,  if the set $\mathcal{F}$$\triangleq$$(0, \hat{f})$ is non-empty (where $\hat{f}$$\ll$$\frac{\sigma}{2\pi\epsilon}$), there exists a unique optimal frequency $f^* \approx {f^{*}_0}^+$ within the set $\mathcal{F}$$=$$(0, \hat{f})$.
	\end{corollary}
	
	Notably,  for most practical underground scenarios (e.g., soil, rock), \( \sigma \) typically ranges from \( 10^{-4} \) to \( 1 \, \text{S/m} \), while \( \epsilon \)  remains at a relatively low order of magnitude. Few materials have a relative permittivity $\epsilon_{\rm r}$ exceeding 300. This makes the value of \( \frac{\sigma}{2\pi\epsilon} \) sufficiently large,  further contributing to a broad upper bound for \( \hat{f} \). This endows \( \mathcal{F} \) with a larger range to accommodate  \( f^* \). Is this case, the  lowest frequency  in the neighborhood of the second optimal point $\infty$  that outperforms  $f^*$  is  extremely high. 
	Hence, a frequency slightly higher than $f_0^*$  can be selected  in practice.
	
	Additionally, according to~\eqref{eqn_f0star1}, $f^*_0\!\propto\!\sigma^{-\frac{3}{5}}$ implies that underground material properties exert a significant influence on the optimal frequency. Thus, the empirical use of 10 kHz is only suitable for limited scenarios. For example, we need to use a lower working frequency for the underground space with wet soil than that with dry soil for a larger MIC distance. In this case, as $\sigma$ and $\hat{f}$ (where $\hat{f}$$\propto$$\frac{\sigma}{2\pi\epsilon}$) increase and given the uniqueness  of $f^*$ in $\mathcal{F}$, both the optimal range (see Remark \ref{rmk_dstsigma}) and the range at the empirical frequency of 10 kHz decrease. Thus, this empirical frequency is no longer a local maximum.
	
	However, \eqref{eqn_f0star1} is overly complex for engineering application.  As the BER = $10^{-3}$ (corresponding to an SNR  $\Upsilon_{\rm th}\approx4.778$) generally indicates reliable communication and $f_0^*$ is derived under $\lim\limits_{\epsilon f\rightarrow0}\frac{\sigma}{2\pi \epsilon f}$$=$$\infty$ (see Appendix \ref{sect_proof}), ~\eqref{eqn_f0star1} can be approximated effectively by solving 	$f_0^* \approx 5.113  (\tfrac{\Upsilon^*_{\backslash f}}{4.778})^{-\frac{1}{5}} \cdot (\tfrac{1}{\delta^2_{\rm u} f_0^*})^{-\frac{3}{5}}$, i.e.,
	\begin{equation}\label{eqn_f0appr}
		\begin{aligned}
			f_0^* &\approx\tfrac{ 6.990}{\sqrt[5]{(\pi \mu \sigma)^{3}\Upsilon^*_{\backslash f}}}\approx 129.2\cdot\tfrac{ \delta^3_u}{\sqrt{\Upsilon^*_{\backslash f}}},
		\end{aligned}
	\end{equation}
	where $\delta_{\rm u}$ is the skin depth of medium, which can be more readily obtained than  $\pi\mu\sigma$.

\vspace{-0.9em}

\subsection{Effect of Underground Materials Conductivity}\label{sectsub_eoum}

Let $t$$=$$2\pi\epsilon f$.  We obtain the derivative of the effective MIC range w.r.t.  the conductivity $\sigma_{}$ of the underground material:

\begin{equation}\label{eqn_ddsdsigma}
	\begin{aligned}
		d_{}'(\sigma_{}) \!=\!\tfrac{3 \mathbb{W}_f^{\!2} \sigma}{4 \sqrt{\pi}\, \sqrt{2 \mu f \left(
				\frac{\sqrt{t^2 + \sigma^2}-\frac{t}{2}}{2}
				\right)}\, \sqrt{t^2 + \sigma^2}\, \left(
			\frac{t}{2} - \frac{\sqrt{t^2 + \sigma^2}}{2}
			\right) \left(1 + \mathbb{W}_f\right)}.
	\end{aligned}
\end{equation}
Since only the factor $\left(
\frac{t}{2} - \frac{\sqrt{t^2 + \sigma^2}}{2}
\right)$  in the denominator of~\eqref{eqn_ddsdsigma} is negative, $d_{}'(\sigma_{}) < 0$ is satisfied for all $\sigma_{}$. Therefore, we arrive at the following conclusion.
\begin{remark}\label{rmk_dstsigma}
	The effective range of TTE MIC decreases with the increase of the conductivity of underground media.
\end{remark}

\vspace{-1.4em}
\section{Simulation and Numerical Results}\label{sect_sim}

This section first verifies the derived effective MIC range, then analyzes its dependence on frequency, transmit power, coil size and medium conductivity via simulation, and finally discusses feasible range enhancement methods.

The underground MIC environments are similar to our previous works~\cite{Ma2019Effect,Ma2019Antenna}, which also use VLF-LA methods. Particularly, we assume the underground space is filled with homogeneous media whose permeability $\mu_{}$ and permittivity $\epsilon_{}$ are consistent with~\cite{kisseleff2013channel,Ma2024Fast}, i.e., $\mu=4\pi \times 10^{-7}$ H$\cdot$m$^{-1}$ and $\epsilon_{}= 6.978 \cdot 10^{-11}$ F/m.  We assume that the number of transmit coil turns is $n_S=15$, and the transmit coil radius is $a_S=0.6$ m (adapting to a vehicle size). The number of receive coil turns is $n_D=30$, receive coil radius is $a_D=0.4$ m. The unit length resistance of the antenna is $\rho_w=0.0166$ $\Omega$/m. We set the transmitting  PSD $P_S = 5$ W / 450 Hz.  The ambient noise PSD is assumed to be $P_N=-103$ dBm / 2000 Hz~\cite{Sun2010Magnetic}. As in~\cite{Ma2019Effect}, we set the threshold $\Upsilon_{\rm th}=4.7748$; i.e., the receive SNR when the BER at the receiving node reaches $10^{-3}$~\cite{Ma2019Effect}. 

\vspace{-1.0em}

\subsection{Effective MIC range}\label{sectsub_emr}

\begin{figure}[t]
	\centering
	%\color{mjColorRevision}
	\includegraphics[width=0.75\linewidth, height=0.45\linewidth]{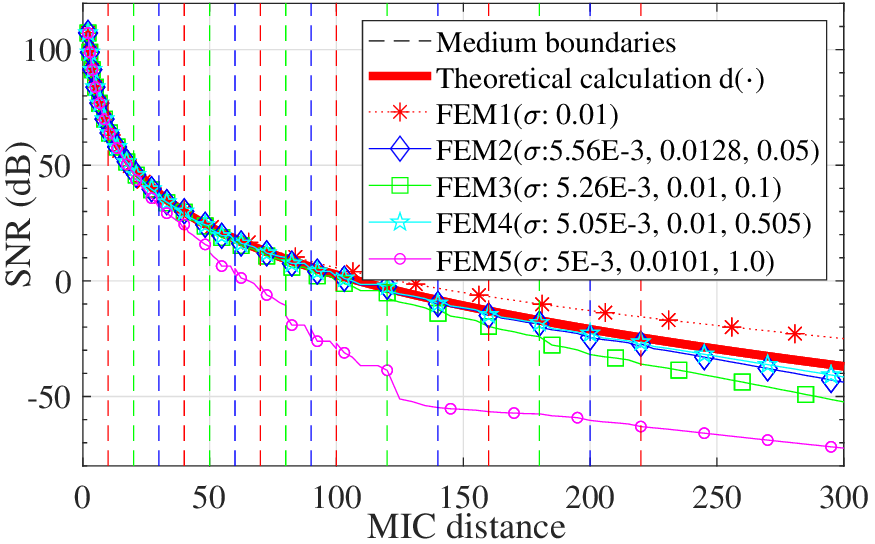}
	\vspace{-1.0em}
	\caption{ Finite element method (FEM) validation for the MI effective range under single- and multi-layered conductive media. Red, green, and blue vertical dashed  lines denote medium boundaries with conductivities $\sigma_1$, $\sigma_2$, and $\sigma_3$, respectively. The  curves FEM1-FEM5 ($\sigma$: $\sigma_1,\sigma_2,\sigma_3$) are  obtained from COMSOL, while the red thick solid line (for a single-layered $\sigma=0.01$ S/m) is calculated via~\eqref{eqn_dsd} with  $\Upsilon_{\rm th}$$=$SNR (in dB) at 1000 Hz.}
	\label{fig_fem}
	\vspace{-1.0em}
\end{figure}

\begin{figure}[t]
	\centering
	%\color{mjColorRevision}
	\includegraphics[width=1\linewidth,  height=0.5\linewidth]{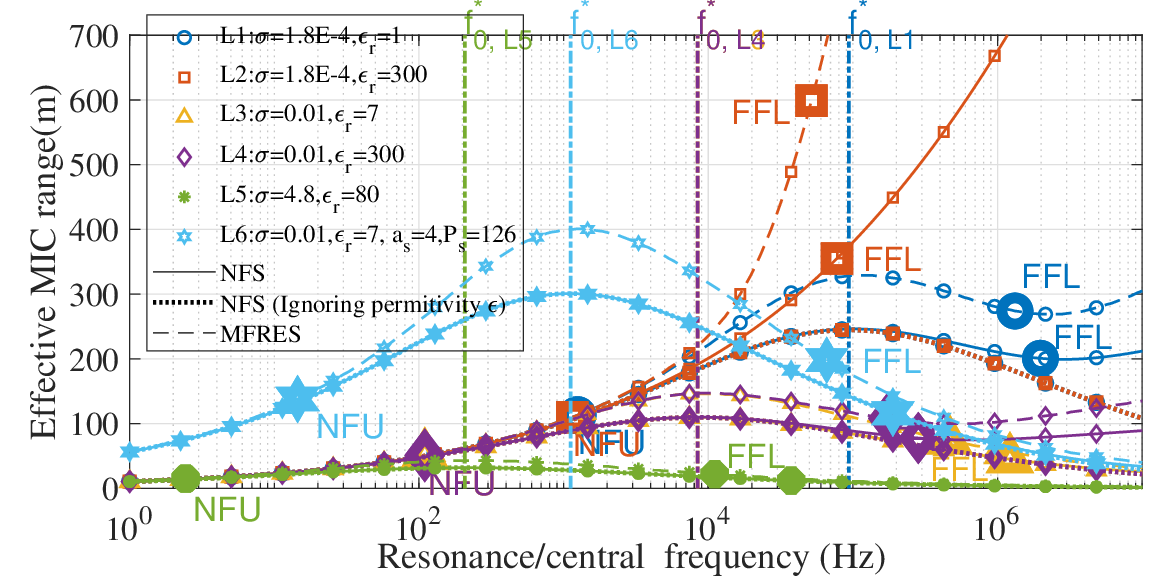}
	\caption{Validation  of Proposition \ref{prop_fstar} and Corollary \ref{cor_fstar};
		NFS: near-field solution~\eqref{eqn_dsd}; MFRES: numerical solution of~\eqref{eqn_dsdeq}; NFU/FFL: near-field upper/far-field lower bounds; vertical dash-dot lines $f^*_{0, {\rm L}(\cdot)}$ are calculated from~\eqref{eqn_f0star1};  $\Upsilon_{\rm th, L6}$$=$$0.125$ (SNR threshold for L6).}% Datong experimental reference distance is from\cite{Zhang2017Connectivity} for reference only.}}
		\label{fig_dsdfreq}
		\vspace{-1.2em}
\end{figure}

To validate~\eqref{eqn_dsd} for the effective MIC range, a comparison between the FEM simulation (FEM1) and theoretical calculation (thick red line) is  presented in Fig.   \ref{fig_fem}. It is evident that the two sets of results exhibit good agreement, especially in the shorter distance range where the two curves almost overlap.    
As the distance increases, the discrepancy between the FEM  and the theoretical calculation curves becomes more pronounced.  This is attributed to the fact that the effect of the radiation field, which is negligible in the near-field region but gradually dominates in the far-field, begins to manifest. Such a comparison  demonstrates the accuracy and reliability of~\eqref{eqn_dsd} 
in predicting the MIC effective range in near-field regime.

\vspace{-0.8em}

\subsection{Frequency Optimization}\label{sectsub_fo}
\vspace{-0.2em}

Fig.~\ref{fig_dsdfreq} plots the analytical effective MIC range  versus the frequency, along with the optimal frequency. The solid lines are calculated from~\eqref{eqn_dsd}; the dotted lines are calculated from~\eqref{eqn_dsd} with $\epsilon\approx0^+$, consistent with underground media configurations in \cite{Zhang2017Connectivity}, and the dashed lines represent  the numerical solutions of MFRE with the radiation field considered. We observe: 1) all peaks of the solid lines are slightly right-shifted relative to corresponding vertical dash-dot lines  $f^*_{0,{\rm L}(\cdot)}$, validating the equality condition in Proposition \ref{prop_fstar} and Corollary \ref{cor_fstar}; 2) for each curve pair, the frequency corresponding to the peak of the dashed line is slightly right-shifted relative to corresponding dash-dot line $f^*_{0,{\rm L}(\cdot)}$, implying that the radiation field has a negligible impact on the first optimal frequency obtained from \eqref{eqn_f0star1};  
3) all peaks lie within the Fresnel region (segments between NFU and FFL), as analyzed in Section \ref{sectsub_of}; 4) in the Fresnel region, the radiation field can offer additional MIC range extension (e.g., L6: up to about 33\%);  
5) there also exists an extreme medium configuration L2 with $\mathcal{F}$$=$$\emptyset$. In this case, $f^*$ does not exist and $d_{} (f)$ increases monotonically with $f$.  However, few scenarios conform to this configuration since the high permittivity is caused by high moisture content, and high moisture can enhance the conductivity;  6) for the same curve,  when the frequency deviates significantly from  $f^*$,  the MIC range decreases by severalfold;  
and 7) there exists a second locally optimal solution. Under the typical underground/undersea configurations of L1, L3, L5, and L6, the second  optimal frequency  should be set to a value close to infinity to achieve better performance than the first one $f^*$.

\begin{figure}[t]
\centering{\includegraphics[height=1.4in]{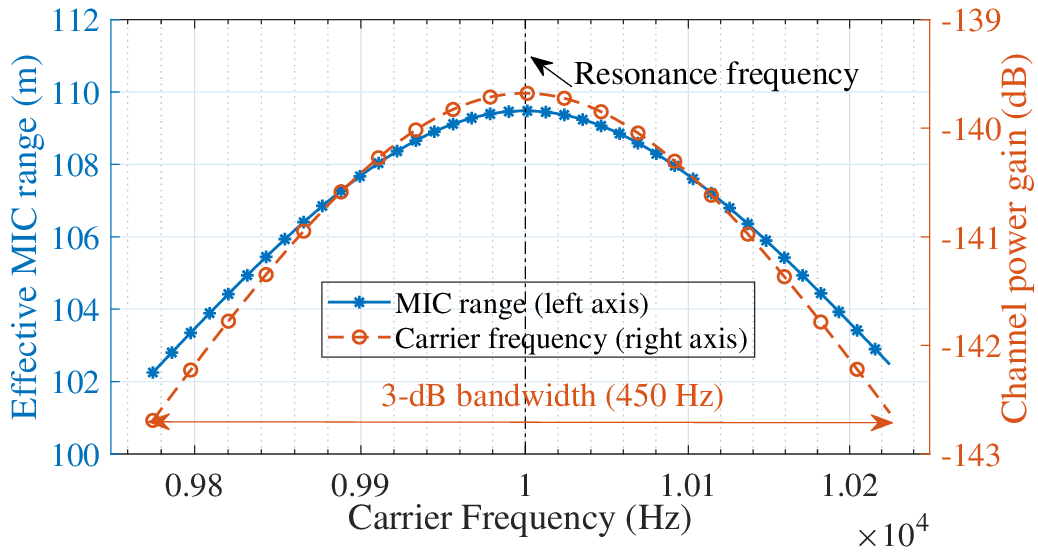}}
\caption{Effective MIC range (left axis) and channel power gain (right axis) versus carry frequency in a 3-dB bandwidth.}\label{fig_freqdsdband}
\vspace{-1.0em}
\end{figure}

Fig.~\ref{fig_freqdsdband} depicts the effective MIC range  versus carrier frequency over the 3-dB bandwidth. The system achieves its maximum MIC range at the designed resonance frequency (10 kHz). Within this bandwidth, the MIC range  decreases by only about 7\% (from 109.5 m to 102 m) at the band edges, demonstrating  robustness to frequency offsets and validating the feasibility of multi-frequency operation within a resonance band. This is due to $d \propto \mathbb{W}(\Upsilon_S^{\frac{1}{6}})$ (see~\eqref{eqn_dsd}), making the MIC range nearly insensitive to variations in received power across the bandwidth.

\vspace{-1.1em}
\subsection{Influence of  Underground Medium}\label{sect_subium}
We simulate the effect of medium conductivity on the MIC range (Fig. \ref{fig_dsdsigma}), and evaluate the deviation from the closed-form solution \eqref{eqn_dsd} in multi-layered media (Fig. \ref{fig_fem}).
\subsubsection{Medium conductivity}\label{sect_subsubmc}
\begin{figure}[t]
\centering{\includegraphics[height=1.6in]{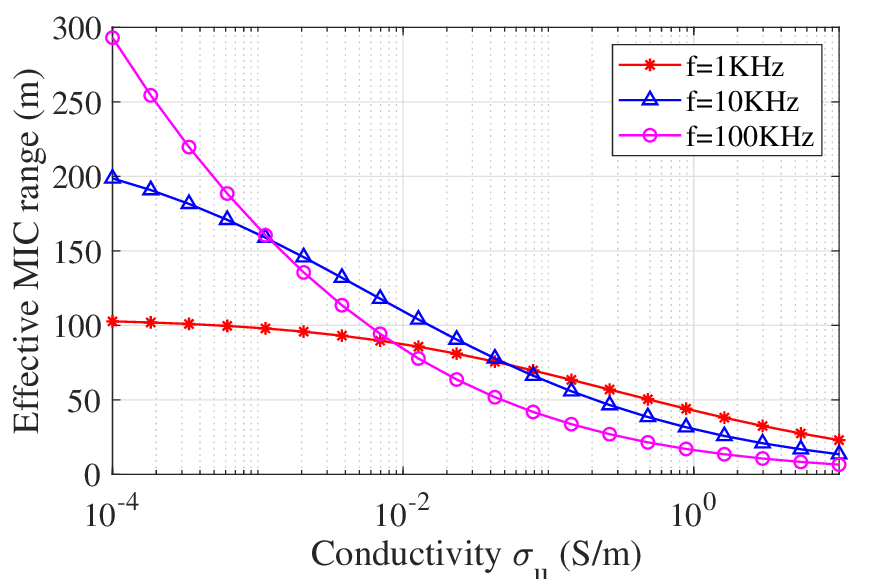}}
\caption{{The effects of material conductivity $\sigma_{}$  on the communication range.}}\label{fig_dsdsigma}
\vspace{-1.3em}
\end{figure}

As shown in Fig.~\ref{fig_dsdsigma}, the effective communication range of the TTE MIC decreases remarkably as the conductivity increases. This validates Remark \ref{rmk_dstsigma} in the analysis.  More importantly, Fig.~\ref{fig_dsdsigma} illustrates that the influence of the underground medium is significantly intensified under high carrier frequencies.
\subsubsection{Multi-layered medium conductivity} \label{sect_subsubmmc}
We evaluate the deviation of the closed-form solution~\eqref{eqn_dsd} via FEM simulations. We consider a multi-layered soil medium within a depth of 0--220 m. Its region from 0 to 100 m is divided into layers of 10 m each, while the region from 100 to 220 m is divided into layers of 20 m each. From the shallowest layer to the deepest,
the  conductivity takes values in a cyclic sequence of $\sigma_1$, $\sigma_2$, and $\sigma_3$ ($\sigma_1$$\leq$$\sigma_2$$\leq$$\sigma_3$), whose series equivalent conductivity is approximately 0.01 S/m, as shown in Fig. \ref{fig_fem}. It can be observed that when the distance is less than 50 m, all curves are in close agreement. When the distance exceeds 50 m and $\sigma_3$$<$$100\sigma_1$ (FEM1--FEM4), the discrepancy between the curves remains small. When the ratio $\sigma_3/\sigma_1$$>$$250$ and $\sigma_3$$>$$1$ (FEM5, with high conductivity), the FEM results deviate significantly from \eqref{eqn_dsd}. These deviations arise from the combined effects of the radiation field and inhomogeneous media.
Thus, the homogeneous model  achieves acceptable accuracy for typical TTE  scenarios with moderate conductivity variations.

\vspace{-1.0em}
\subsection{Comparison of effective MIC range optimization}
\begin{figure}[t]
\centering{\includegraphics[height=1.7in]{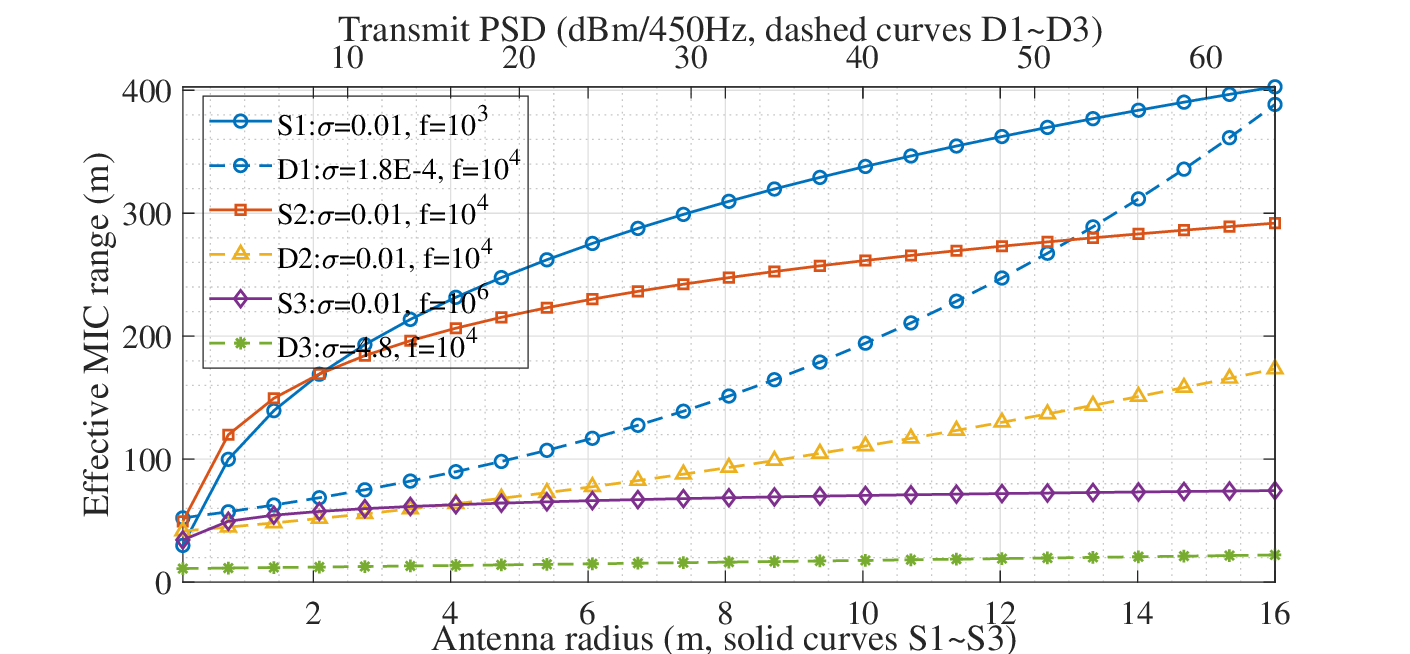}}
\caption{{Simulations of other optimization approaches (Transmit PSD and coil size) for effective MIC range.} }
\label{fig_other} %% label for entire figure
\vspace{-1.5em}
\end{figure}

We  discuss the potential optimizations of transmit power and antenna designs, as well as the frequency optimization, for achieving a longer effective MIC range through numerical simulations. As exhibited in~\eqref{eqn_dsd}, the effective MIC range increases as the transmit power increases and coil size.  Thus, as observed in Fig.~\ref{fig_other}, when the transmit power is increased from 0.001 W to 1,000 W, the MIC range increases by only several times. This suggests that increasing the transmit power to enhance the MIC range is energetically inefficient.  In contrast, although adjusting the carrier frequency $f$ to the optimal frequency $f^*$ can also vary by thousands of times in terms of frequency, varying the frequency is energy-efficient and does not incur significant software or hardware costs.

As also noticed in Fig.~\ref{fig_other},  increasing the coil radius can initially expand the MIC range significantly. The gains diminish as the radius continues to increase, and this tendency is more pronounced at higher frequencies. Thus, there exists an optimal coil size that maximizes the MIC range without undue enlargement. This simulation further validates the rationality of VLF-LA for TTE MIC. 
To this end, optimizing the coil radius and carrier frequency are two effective strategies for enhancing the MIC range. However, enlarging the coil radius significantly increases the size of the MI device, leading to higher deployment and device costs.

%\vspace{-0.4em}
\section{Conclusion}\label{sect concl}

In this paper, we derived closed-form expressions for the effective MIC range and the optimal frequency that maximizes the effective MIC range, accounting for the conductivity and permittivity of underground materials. FEM simulation validates the MIC range. Through these expressions and simulation results, we compared various MIC range enhancement strategies, including optimizing transmit power, antenna radius, and carrier frequency. Among these, optimizing the antenna radius and carrier frequency proved most effective. Moreover, while optimizing the carrier frequency can enhance the MIC range without additional software or hardware costs, increasing the antenna radius, though effective, may significantly raise device size and associated costs.
\vspace{-0.0em}

\appendices

\section{Proof of Proposition \ref{prop_fstar}}\label{sect_proof}

Let $\mathcal{K}$$=$$\sqrt{\frac{4 \pi^{2} \epsilon^{2} f^{2}+\sigma^{2}}{f^{2}}}$ and $\mathbb{W}_f$$=$$\mathbb{W}( \mathcal{A}(f) )$, where $\mathcal{A}(f)$$=$$\frac{\sqrt{\mu}}{3} \sqrt{\pi  f \sqrt{4\pi^2 \epsilon^2 f^2 + \sigma^2} - 2\pi^2 \epsilon f^2}  ( \frac{\Upsilon_{\backslash f}}{\Upsilon_{\rm th}}  f^2)^{\frac{1}{6}}$,  $\Upsilon_{\backslash f}=\frac{\Upsilon_S(f)}{f^2}$. The optimal frequency $f^*$ can be obtained by solving $d_{}'(f)=0$ where $d_{}'(f)$ denotes the derivative of $d_{}(f)$ w.r.t. $f$. We have
\begin{equation}\label{eqn_ddsd}
	\begin{aligned}
		0 &= \mathbb{W}_f \big[ 6\left(\pi \mathcal{K}\epsilon f^2 - 2\pi^2\epsilon^2f^2 -\tfrac{\sigma^2}{4}\right)\mathbb{W}_f \\
		&\quad + \tfrac{2\pi^2\epsilon^2f^2}{3} - \tfrac{\mathcal{K}\pi\epsilon f^2}{3} + \tfrac{\sigma^2}{6} \big] \\
		&\quad \big/ \big[ \mathcal{K}f^4\sqrt{\pi}\sqrt{\mu\mathcal{K}-2\mu\epsilon\pi} 
		\quad (1+\mathbb{W}_f) (\mathcal{K}-2\epsilon\pi) \big]
	\end{aligned}
\end{equation}
Given that all variables in $d_{}(f)$ are positive, the Lambert-W function $\mathbb{W}_f$ takes a positive value. Thus, \eqref{eqn_ddsd} can be rearranged into
\begin{equation}
	\begin{aligned}
		\mathbb{W}(\mathcal{A}(f))	\!=\!\tfrac{1}{3} \!+\! \tfrac{-\sigma^2}{12 \left( \pi \epsilon f \sqrt{4\pi^2 \epsilon^2 f^2 + \sigma^2} - 2\pi^2 \epsilon^2 f^2 - \frac{\sigma^2}{4} \right)}.
	\end{aligned}
\end{equation}
Subsequently, we analyze the number of intersections between $\mathbb{W}(\mathcal{A}(f))$ and $\mathcal{R}(f)$$=$$\tfrac{1}{3}$$+$$\tfrac{-\sigma^2}{12 \left( \pi \epsilon f \sqrt{4\pi^2 \epsilon^2 f^2 + \sigma^2} - 2\pi^2 \epsilon^2 f^2 - \frac{\sigma^2}{4} \right)}$.	 Let $\mathcal{A}(f)$$=$$C{\mathcal{B}(f)} f^\frac{1}{3} $ where $\mathcal{C}$$=$$\frac{\sqrt{\mu}}{3}\frac{\Upsilon_{s}}{\Upsilon_{\rm th}}$, $\mathcal{B}(f)$$=$$\sqrt{\mathcal{S}(f)}$ $\mathcal{S}(f)$$=$${\pi  f \sqrt{4\pi^2 \epsilon^2 f^2 + \sigma^2} - 2\pi^2 \epsilon f^2}$$=$$\frac{\pi\sigma^2f}{\sqrt{4\pi^2\epsilon^2f^2 + \sigma^2}+2\pi\epsilon f}$.  Obviously, both $\mathcal{A}(f)$ and its derivative $\mathcal{A}'(f)$ are greater than 0 due to the derivative $\mathcal{B}'(f)$$=$$\frac{\pi \sigma^{4}}{2 \sqrt{\mathcal{S}(f)}  \sqrt{4 \pi^{2} \epsilon^{2} f^{2} + \sigma^{2}( \sqrt{4\pi^2\epsilon^2f^2 + \sigma^2}+2\pi\epsilon f )^{2}}}$$>$$0$. Thus, the function $\mathbb{W}(\mathcal{A}(f))$ is monotonically increasing.
%It can be observed that the numerator of $\mathcal{B}'(f)$ increases at a faster rate than its denominator. Hence, 
As the second derivative  $\mathcal{B}''(f)$$<$$0$, and the second derivative $\mathcal{A}''(f)$$=$$\mathcal{C}((f^\frac{1}{3})\mathcal{B}''$$+$$(f^\frac{1}{3})''\mathcal{B}$$+$$2(f^\frac{1}{3})'\mathcal{B}')$$<$$0$ holds. Thus, the function $\mathcal{A}(f)$ is  monotonically increasing and concave downward within $f \in (0, \infty)$.

Similarly, let $\mathcal{R}(f)$$=$$\frac{1}{3}$$-$$\frac{\sigma^2}{12}\frac{1}{\mathcal{D}(f)}$  where $\mathcal{D}(f)$$=$$\pi \epsilon f \sqrt{4\pi^2 \epsilon^2 f^2 + \sigma^2}$$-$$2\pi^2 \epsilon^2 f^2$$-$$\frac{\sigma^2}{4}$. {Since the denominator $\mathcal{D}(f)$ is monotonically increasing and concave  within $f$$\in$$(0,\infty)$ (i.e., $\mathcal{D}''$$=$$4\pi^2\epsilon^2 [\frac{\pi\epsilon f \left(4\pi^2\epsilon^2f^2 + 3\sigma^2\right)}{(4\pi^2\epsilon^2f^2 + \sigma^2)^{3/2}}$$-$$1]$$<$$0$), 
	the function
	$\mathcal{R}(f)$ is monotonically increasing and convex when $\mathcal{R}(f)$$>$$\frac{1}{3}$ (due to $\mathcal{R}''$$=$$\frac{\sigma^2}{12} \frac{2 \left( \mathcal{D}' \right)^2-\mathcal{D} \mathcal{D}''}{\left( \mathcal{D} \right)^3}$$>$$0$ when $\mathcal{D}>0$).
	
	Given the convexity,  concavity and monotonicity of $\mathbb{W}(\mathcal{A}(f))$ and  $\mathcal{R}(f)$, together with the limit conditions $\lim\limits_{f\rightarrow\infty}\mathbb{W}(\mathcal{A}(f))>\frac{1}{3}$ and $\lim\limits_{f\rightarrow\infty}\mathcal{R}(f) = \frac{1}{3}$, the number of intersections between $\mathbb{W}(\mathcal{A}(f))$ and $\mathcal{R}(f)$ is at most 2. Therefore, $d_{}(f)$ has at most two stationary points, denoted as $f^*_L$ and $f^*_H$ ($f^*_L \leq f^*_H$). Given $d_{}(f) > 0$ and $\lim\limits_{f\rightarrow0}d_{}(f)$$=$$0$, $d_{}(f^*_L)$ and $d_{{sd}}(f^*_H)$ correspond to the maximum and minimum values, respectively. This indicates that there exist at most two locally optimal frequencies, $f^*_L$ and $\infty$ (right endpoint).
	
	In particular, if $\epsilon f$$=$$0$,  $\alpha$$=$$\sqrt{{\pi f \mu \sigma}}$ holds. Under such a scenario,
	$\mathcal{R}(f)$$=$$\frac{2}{3}$ is a constant, and $\mathbb{W}(\mathcal{A}(f))$ has a unique intersection with $ \mathcal{R}(f)$. This indicates the existence of a unique stationary point $ f_L^*$$=$$ 2^{6/5}  \mathrm{e}^{4/5} ( (\tfrac{\Upsilon^*_{\backslash f}}{\Upsilon_{\rm th}})^{1/3}\pi\mu_u\sigma)^{-3/5}$$\triangleq$$f_0^*$. Thus,  the sufficient condition for the equality in \eqref{eqn_fstar} holds.
	
	Notice that $\mathbb{W}(\mathcal{A}(f, \epsilon f$$>$$0))$$<$$\mathbb{W}(\mathcal{A}(f, \epsilon f$$=$$0))$ and $\mathcal{R}(f, \epsilon f$$>$$0)$$>$$\mathcal{R}(f, \epsilon f$$=$$0)$  both hold.  These inequalities indicate that when  $\epsilon f$$\neq$$0$, the intersections of $\mathbb{W}(\mathcal{A}(f))$ and $\mathcal{R}(f)$ move rightward from $f_0^*$, i.e., $f_L^*$$\geq$$f_0^*$.  On the other hand, combined with this rightward shift property, these inequalities also imply that  the condition $\epsilon f$$\neq$$0$ cannot hold when  $f_L^*$$=$$f_0^*$ due to $f^*_L$$<$$f^*_H$. Thus, the necessary condition for the equality in \eqref{eqn_fstar} is also satisfied.
	
	Further, as $\epsilon f$ increases beyond a threshold, the number of intersections (i.e., stationary points) becomes zero. Since  $\lim\limits_{f\rightarrow0}d_{}(f)$$=$$0$ and $d_{}(f)$$>$$0$ hold, the function $d_{}(f)$  increases monotonically with $f$,  and the right endpoint $\infty$ becomes the  unique locally optimal  frequency. 
	
	This concludes the proof.

% %%%add by hlma
\bibliographystyle{IEEEtranet} % 
\bibliography{IEEEabrv, MIref}

@STRING{IEEE_J_VT         = "{IEEE} Trans. Veh. Technol."}

@STRING{IEEE_J_COML       = "{IEEE} Commun. Lett."}

@STRING{IEEE_J_COM        = "{IEEE} Trans. Commun."}

@STRING{IEEE_J_WCOM       = "{IEEE} Trans. Wireless Commun."}

@STRING{IEEE_J_AP         = "{IEEE} Trans. Antennas Propag."}

@STRING{IEEE_O_CSTO       = "{IEEE} Commun. Surveys Tuts."}

@Article{Sun2010Magnetic,
  author  = {Zhi Sun and Ian F. Akyildiz},
  title   = {Magnetic Induction Communications for Wireless Underground Sensor Networks},
  journal =IEEE_J_AP,
  year    = {2010},
  volume  = {58},
  number  = {7},
  pages   = {2426-2435},
  month   = jul,
}

@book{Rappaport1996wireless,
  title={Wireless Communications: Principles and Practice},
  author={Rappaport, Theodore S.},
  publisher={Prentice Hall},
  year={1995},
  ISBN={0-7803-1167-1},
  address={Upper Saddle River, NJ, US},
}

@InProceedings{zhang2014cooperative,
  author    = {Zhang, Zhengqing and Liu, Erwu and Zheng, Xiaojun and Jian, Yuhui and Wang, Dong and Liu, Dong},
  title     = {Cooperative magnetic induction based through-the-earth communication},
  booktitle = {IEEE/CIC ICCC},
  year      = {2014},
  pages     = {653--657},
  address   = {Shanghai, China},
  month     = oct,
}

@Article{Sun2013Increasing,
  author   = {Sun, Zhi and Akyildiz, Ian F. and Kisseleff, Steven and Gerstacker, Wolfgang},
  journal  = IEEE_J_COM,
  title    = {Increasing the Capacity of Magnetic Induction Communications in {RF}-Challenged Environments},
  year     = {2013},
  number   = {9},
  pages    = {3943-3952},
  volume   = {61},
  doi      = {10.1109/TCOMM.2013.071813.120600},
}

@Article{Zhang2017Connectivity,
  author  = {Zhengqing Zhang and Erwu Liu and Xinyu Qu and Rui Wang and Honglei Ma and Zhi Sun},
  title   = {Connectivity of Magnetic Induction-based Ad Hoc Networks},
  journal = {IEEE Transactions on Wireless Communications},
  year    = {2017},
  volume  = {16},
  number  = {7},
  pages   = {4181-- 4191},
  month   = apr,
  doi     = {10.1109/TWC.2017.2693184},
}

@InProceedings{kisseleff2013channel,
  author    = {Kisseleff, Steven and Gerstacker, W and Schober, Robert and Sun, Zhi and Akyildiz, Ian F},
  title     = {Channel capacity of magnetic induction based wireless underground sensor networks under practical constraints},
  booktitle = {IEEE WCNC},
  year      = {2013},
  pages     = {2603--2608},
  address   = {Shanghai, China},
  month     = apr,
}

@Article{Ma2019Effect,
  author   = {H. {Ma} and E. {Liu} and R. {Wang} and X. {Qu}},
  title    = {Effect of Antenna Deployment on Achievable Rate in Cooperative Magnetic Induction Communication},
  journal  = IEEE_J_COML,
  year     = {2019},
  volume   = {23},
  number   = {10},
  pages    = {1748--1752},
  month    = oct,
  doi      = {10.1109/LCOMM.2019.2929790},
}

@Article{Ma2019Antenna,
  author   = {H. {Ma} and E. {Liu} and R. {Wang} and X. {Yin} and X. {Qu} and Z. {Xu} and B. {Li}},
  title    = {Antenna Optimization for Decode-and-Forward Relay in Magnetic Induction Communications},
  journal  = IEEE_J_VT,
  year     = {2019},
  pages={3449-3453},
  volume={69},
  number={3},
  issn     = {1939-9359},
  doi      = {10.1109/TVT.2019.2963357},
}

@ARTICLE{Li2022Optimal,
  author={Li, Zhangyu and Sun, Zhi},
  journal=IEEE_J_AP, 
  title={Optimal Active and Reconfigurable Meta-Sphere Design for Metamaterial-Enhanced Magnetic Induction Communications}, 
  year={2022},
  volume={70},
  number={9},
  pages={8148-8163},
  doi={10.1109/TAP.2022.3164164}}

@Article{Guo2021Joint,
  author    = {Hongzhi Guo and Zhi Sun and Pu Wang},
  journal   = IEEE_J_VT,
  title     = {Joint Design of Communication, Wireless Energy Transfer, and Control for Swarm Autonomous Underwater Vehicles},
  year      = {2021},
  issn      = {0018-9545},
  number    = {2},
  pages     = {1821-1835},
  volume    = {70},
  doi       = {10.1109/tvt.2021.3053456},
  publisher = {Institute of Electrical and Electronics Engineers (IEEE)},
}

@Article{Guo2017Multiple,
  author   = {Guo, Hongzhi and Sun, Zhi and Wang, Pu},
  journal  = IEEE_J_VT,
  title    = {Multiple Frequency Band Channel Modeling and Analysis for Magnetic Induction Communication in Practical Underwater Environments},
  year     = {2017},
  number   = {8},
  pages    = {6619-6632},
  volume   = {66},
  doi      = {10.1109/TVT.2017.2664099},
}

@Article{Ma2024Fast,
  author   = {Ma, Honglei and Liu, Erwu and Fang, Zhijun and Wang, Rui and Gao, Yongbin and Yu, Wenjun and Zhang, Dongming},
  journal  = IEEE_J_WCOM,
  title    = {Fast-Fading Channel and Power Optimization of the Magnetic Inductive Cellular Network},
  year     = {2024},
  number   = {10},
  pages    = {15096-15111},
  volume   = {23},
  doi      = {10.1109/TWC.2024.3425473},
}

@InProceedings{Zhou2017Maximum,
  author    = {Jiaxin Zhou and Jianyong Chen},
  booktitle = {IEEE CCWC},
  title     = {Maximum distance estimation of far-field model for underwater magnetic field communication},
  year      = {2017},
  address   = {Las Vegas, NV, USA},
  month     = jan,
  pages     = {1-5},
  doi       = {10.1109/CCWC.2017.7868371},
}

@Book{Liu2024Magnetic,
  author    = {Liu, Erwu and Sun, Zhi and Wang, Rui and Guo, Hongzhi},
  publisher = {Cambridge University Press},
  title     = {Magnetic Communications: Theory and Techniques},
  year      = {2024},
  month     = jan,
  place     = {Cambridge},
}

@InProceedings{Zhang2024Cooperative,
  author    = {Zhang, Yixin},
  booktitle = {IEEE WCNC},
  title     = {Cooperative Magnetic Induction Communications: Performance Analysis and Power-Location Optimization},
  year      = {2024},
  address   = {Dubai, United Arab Emirates},
  pages     = {1-6},
  doi       = {10.1109/WCNC57260.2024.10570545},
}

@Article{Sun2024Improved,
  author   = {Sun, Qian and Wang, Hao and Liu, Wa and Zou, Junjing and Ye, Fang and Li, Yibing},
  journal  = IEEE_J_VT,
  title    = {An Improved Stereo Visual-Inertial SLAM Algorithm Based on Point-and-Line Features for Subterranean Environments},
  year     = {2025},
  number   = {3},
  pages    = {3925-3940},
  volume   = {74},
  doi      = {10.1109/TVT.2024.3492388},
}

@Article{Jenkins2023Wearable,
  author   = {Jenkins, Connor B. and Kiourti, Asimina},
  journal  = IEEE_J_AP,
  title    = {Wearable Dual-Layer Planar Magnetoinductive Waveguide for Wireless Body Area Networks},
  year     = {2023},
  number   = {8},
  pages    = {6893-6905},
  volume   = {71},
  doi      = {10.1109/TAP.2023.3286042},
}

@ARTICLE{Dong2020Velocity,
  author={Dong, Longjun and Sun, Daoyuan and Han, Guangjie and Li, Xibing and Hu, Qingchun and Shu, Lei},
  journal=IEEE_J_VT, 
  title={Velocity-Free Localization of Autonomous Driverless Vehicles in Underground Intelligent Mines}, 
  year={2020},
  volume={69},
  number={9},
  pages={9292-9303},
  doi={10.1109/TVT.2020.2970842}}

@ARTICLE{Ko2024Field,
  author={Ko, Kyeongjun and Kim, Youngju and Yoon, Yongki and Park, Sungsoo},
  journal=IEEE_J_VT, 
  title={Field Verification of Wireless Cellular Communication-Based Subway Train Localization}, 
  year={2024},
  volume={73},
  number={6},
  pages={7681-7692},
  doi={10.1109/TVT.2024.3355888}}

@Article{Ma2026Through,
  author   = {Ma, Honglei and Liu, Erwu and Ni, Wei and Fang, Zhijun and Wang, Rui and Gao, Yongbin and Niyato, Dusit and Hossain, Ekram},
  journal  = IEEE_O_CSTO,
  title    = {Through-the-Earth Magnetic Induction Communication and Networking: A Comprehensive Survey},
  year     = {2026},
  pages    = {2263-2305},
  volume   = {28},
  doi      = {10.1109/COMST.2025.3623258},
}
% %% end hlma

\vspace{11pt}

\newpage

\vfill

\end{document}